# Using the past to constrain the future: how the palaeorecord can improve estimates of global warming

Tamsin L. Edwards*
School of Geographical Sciences, University of Bristol, University Road, Bristol, BS8 1SS, UK
Tel: +44 (0)117 33 17568
Fax: +44 (0)117 928 7878
Email: tamsin.edwards@bristol.ac.uk

Michel Crucifix
Institut d'Astronomie et de Géophysique Georges Lemaître, Université catholique de Louvain, 2 Chemin du Cyclotron, B-1348 Louvain-la-Neuve, Belgium
Tel: +32 10 47 33 00
Fax: +32 10 47 47 22
Email: michel.crucifix@uclouvain.be

Sandy P. Harrison
School of Geographical Sciences, University of Bristol, University Road, Bristol, BS8 1SS, UK
Tel: +44 (0)117 33 17223
Fax: +44 (0)117 928 7878
Email: sandy.harrison@bristol.ac.uk

*Author for correspondence.

**Abstract**

Climate sensitivity is defined as the change in global mean equilibrium temperature after a doubling of atmospheric $CO_2$ concentration and provides a simple measure of global warming. An early estimate of climate sensitivity, 1.5-4.5°C, has changed little subsequently, including the latest assessment by the Intergovernmental Panel on Climate Change.

The persistence of such large uncertainties in this simple measure casts doubt on our understanding of the mechanisms of climate change and our ability to predict the response of the climate system to future perturbations. This has motivated continued attempts to constrain the range with climate data, alone or in conjunction with models. The majority of studies use data from the instrumental period (post-1850) but recent work has made use of information about the large climate changes experienced in the geological past.

In this review, we first outline approaches that estimate climate sensitivity using instrumental climate observations and then summarise attempts to use the record of climate change on geological timescales. We examine the limitations of these studies and suggest ways in which the power of the palaeoclimate record could be better used to reduce uncertainties in our predictions of climate sensitivity.

# Introduction

The concentration of greenhouse gases ($CO_2$, $CH_4$, etc) in the atmosphere has increased substantially during recent decades and is likely to continue increasing in the future. Predictions of how these changes will affect climate are highly uncertain, in part because of uncertainties in how much of the $CO_2$ will be taken up by the ocean, in part because of imperfect knowledge of the terrestrial carbon cycle but most importantly because of uncertainties in the sensitivity of the climate system to change. The effects of increasing greenhouse gas (GHG) concentrations include changes in temperature, precipitation, cloudiness and the frequency of extreme events; changes occur regionally and seasonally and affect short-term climate variability (the chaotic behaviour of climate over timescales of a few years). A simple metric is required to summarise our understanding of future change.

Climate sensitivity is such a metric, defined as the change in average global temperature after atmospheric $CO_2$ concentration is doubled and equilibrium is reached (Schlesinger and Mitchell, 1987). This definition as the equilibrium response distinguishes climate sensitivity from the time-dependent, or 'transient', response of the climate system (see for example Collins et al., 2006; Harris et al., 2006).

The first estimate of climate sensitivity was made over a century ago by Svante Arrhenius who made, in his own words, 'tedious calculations' of the atmospheric temperature change with a doubling of $CO_2$ concentration at various latitudes (Arrhenius, 1896). However, the issue of estimating climate sensitivity was not revisited until the development of atmospheric general circulation models (GCMs) during the 1960s and 1970s (e.g. Manabe and Wetherald 1967, 1975) and the necessary computing power for these models.

Predictions of climate sensitivity from multiple atmospheric GCMs made in the 1970s and summarised in a report by the National Academy of Sciences (NAS), indicated values between 1.5-4.5°C (Charney, 1979). Twenty years later, the IPCC Third Assessment Report presented estimates of climate sensitivity based on more sophisticated coupled ocean-atmosphere GCMs in the range 2-5.1°C, but concluded with the statement that the NAS range "encompasses the estimates from the current models in active use" (IPCC, 2001). The most recent IPCC assessment, published this year, concludes that climate sensitivity is "likely to be in the range 2 to 4.5°C with a best estimate of about 3°C, and is very unlikely to be less than 1.5°C", where 'likely' is defined as greater than about 66% probability and 'unlikely' as less than about 10% probability (IPCC, 2007).

The persistence of such large uncertainties in the estimation of climate sensitivity poses serious problems. Attempts to plan for and adapt to possible future climate change are crucially dependent on knowing the magnitude of that change, which in turn is dependent on the sensitivity of the system. This motivates the current drive to constrain climate sensitivity either by narrowing the range of estimates or by quantifying the probability that climate sensitivity lies within a given range. Better-constrained estimates of climate sensitivity also underpin the reliability of predictions of climate variables other than temperature, changes in climate extremes, and regional climate change.

Most attempts to constrain the range of climate sensitivity have focused on climate changes during the recent past, making use of climate observations, alone or in conjunction with models. However, the changes in global temperature during the historic period (post-1850) are relatively small compared with the change that might be produced by a doubling of $CO_2$. More recently, attempts have been made to constrain climate sensitivity using information about the large climate changes experienced in the geological past, including times when temperature and $CO_2$ concentration were both higher and lower than the present day. In this review, we outline approaches based on historical climate observations since the IPCC Third Assessment Report in 2001, and then summarise attempts to use the record of climate change on geological timescales. We examine the limitations of these studies and suggest ways in which the power of the palaeoclimate record could be better used to reduce uncertainties in our prediction of the future.

**The feedback parameter**

A forcing, broadly speaking, is something that perturbs the radiative equilibrium of the atmosphere: for example, an increase in greenhouse gases (GHGs) that decreases the amount of longwave radiation emitted to space and thus warms the Earth. Other forcings include changes in incoming solar radiation (insolation), ice-sheet extent, and atmospheric levels of sulphate particles (from industrial emissions or volcano eruptions) or black-carbon particles (from industrial emissions or natural fires). A forcing pushes the climate into a warmer or cooler state; this change in state is known as the temperature response. Depending on the size and type of forcing it may take thousands of years for the climate to come into equilibrium. Forcings are measured in Watts per metre squared ($Wm^{-2}$) and can be defined in various ways (Hansen et al., 1997).

The climate's response to a forcing is complicated by the presence of feedbacks, which can amplify or reduce the temperature change. The melting of ice at the poles in response to atmospheric warming produces a positive feedback: the surface albedo (reflectivity) decreases, which decreases the amount of radiation reflected back to space, and so the initial warming is amplified. Some components of the climate system, such as dust or clouds, can produce either positive or negative feedbacks depending on their location and internal characteristics. Low, white clouds have a high albedo so they reflect solar radiation back to space, which has a cooling effect (negative feedback); high, dark clouds trap and re-emit radiation emitted from the Earth, which has a warming effect (positive feedback).

The distinction between forcing and feedback depends on the timescale under consideration, and how it compares with the response times of the different components of the climate system. The response of polar ice sheets to insolation changes, for example, which is slow compared with the response of the atmosphere, can be considered as a feedback on decadal timescales but a forcing on millennial timescales. This allows us to treat the climate system as close to radiative equilibrium on timescales of a few centuries, which is not necessarily the case when long timescale ice sheet dynamics are taken into account.

After a forcing acts to change the radiative balance, and the climate reaches equilibrium at a new temperature, the global mean temperature change ΔT (°C) is related to the forcing Q (Wm$^{-2}$) by:

$$\Delta T = \frac{Q}{\lambda}, \qquad (1)$$

where λ is the feedback parameter (Wm$^{-2}$K$^{-1}$). The simplest way to estimate climate sensitivity is to calculate λ for a known forcing and temperature change ($Q_{known}$, $\Delta T_{known}$), and then use λ with the forcing of doubled CO$_2$ ($Q_{2xCO2}$ = 3.7Wm$^{-2}$) to estimate the temperature change $\Delta T_{2xCO2}$. This makes the assumption that the feedback parameter is the same in both the known and doubled CO$_2$ climates.

**Using the instrumental record**

There are three basic approaches to using modern (post-1850) climate data to estimate climate sensitivity: to infer climate sensitivity directly from observations (using $Q_{known}$, $\Delta T_{known}$ as described in the previous section); to compare model simulations to observations in order to increase confidence in a model's estimate of climate sensitivity; or to weight predictions of climate sensitivity from several different models according to the degree of agreement between the model simulations and the observations.

Climate sensitivity has been inferred by comparing the change in forcing during the instrumental period with observations of atmospheric warming, taking into account the fact that the climate is not at equilibrium (Gregory et al., 2002; Forster and Gregory, 2006). This is a conceptually simple approach, but has practical difficulties. The change in forcing between 1850 and the present day includes changes in solar radiation, atmospheric GHG concentrations, and atmospheric particulate levels including sulphate aerosols. Although the changes in $CO_2$ during this period are well-known, measurements of several of the other forcings only began recently so the change in total forcing must be estimated from a combination of observations and modelling studies. The transient state of the climate is estimated from the rate of heat uptake by the ocean (Gregory et al., 2002), which is quite uncertain, or the net radiative flux imbalance at the top of the atmosphere (Forster and Gregory, 2006), which has only been measured for a short time. A further difficulty is caused by the fact that the global temperature change from 1850 to the present is small (around 0.7°C) and the trend is complicated by natural variability. Because of the large uncertainties in estimating the forcing and ocean heat uptake, the estimate of climate sensitivity (Figure 2, Table 1) by Gregory et al. (2002) has only a lower limit (1.6°C). In a later study, Forster and Gregory (2006) use satellite measurements of the radiative imbalance at the top of the atmosphere and this yields a stronger constraint on climate sensitivity (1-4°C). These estimates based on modern climate observations do not include the albedo forcing due to changes in land cover; it has been estimated that deforestation could have decreased global mean temperature by up to 0.25°C with larger regional changes (Betts, 2001; Brovkin et al., 2006; Davin et al., 2007).

Climate sensitivity has also been estimated by comparing model simulations of the historic period with observations. Models incorporate the major processes that govern climate so, providing these processes are well-represented in the models, they can be used to estimate the impact of each forcing during the historic period and thus the climate sensitivity. The instrumental record can be used for 'model validation', which increases confidence in a model's estimate of climate sensitivity (Barnett et al., 2001; Yokohata et al., 2005), or else to constrain a range of model estimates directly. Ranges of estimates arise because of problems inherent in estimating climate sensitivity from model experiments. Firstly, not all processes and feedbacks are incorporated even in state-of-the-art models. Furthermore, some processes are represented in a simplified fashion (parameterized) and require the specification of parameter values from observations; parameterizations vary from model to model and, in many cases, modern observations yield a range of plausible values for each parameter.

One way to estimate the uncertainties caused by incomplete models and poorly constrained parameters is to run different versions of a given model, and to compare the climate simulated by each version with observations in order to assess which is the most realistic. There are two approaches to creating a group of model versions, or 'ensemble'. The first is to explicitly vary the climate sensitivity of the model, which is usually only possible in simpler models (Andronova and Schlesinger, 2001; Knutti et al., 2002; Harvey and Kaufmann, 2002; Wigley et al., 2005; Frame et al., 2005; Forest et al., 2002, 2006). The second is to vary the parameters that affect the physics in the model, within ranges that are thought to be reasonable (Murphy et al., 2004;

Stainforth et al., 2005; Piani et al., 2005; Knutti et al., 2006). These are referred to as 'climate sensitivity' and 'perturbed physics' ensembles respectively.

Ensembles provide a powerful tool to explore climate sensitivity. Each ensemble member has a different value of climate sensitivity and each simulates the modern climate somewhat differently. The range of climate sensitivity values from the models is expressed as a 'probability distribution function' (pdf) to show which estimates of climate sensitivity are most likely (Figure 1); this first, or 'prior', distribution shows only the predictions of each model and thus reflects the choice of model versions in creating the ensemble (these choices may be referred to as a 'uniform prior' if intended to be neutral, or an 'expert prior' if intended to include a greater degree of opinion: Forest et al., 2002, 2006; Figure 2; Table 1). Each ensemble member is then weighted by its success at simulating the modern climate. The weightings alter the shape of the climate sensitivity pdf, assigning a higher probability to the predictions of the most successful ensemble members and producing the 'posterior' distribution. The posterior distribution is thus made up of the predictions of the ensemble constrained by the climate observations (Figure 1).

Several different instrumental records have been used in the ensemble approach, including observations of present-day climate (Murphy et al., 2004; Stainforth et al., 2005; Piani et al., 2005; Knutti et al., 2006), the long-term warming trends of the atmosphere and ocean in the $19^{th}$ and $20^{th}$ centuries (Andronova and Schlesinger 2001; Knutti et al, 2002; Harvey and Kaufmann, 2002; Frame et al., 2005; Forest et al., 2002, 2006), and observations of cooling after recent volcanic eruptions (Wigley et al. 2005).

Climate sensitivity estimates from ensembles are usually expressed as 5-95% confidence limits (CL), which are the upper and lower limits of the central 90% area of the pdf. Most studies predict an asymmetric climate sensitivity pdf, with a long high tail that would indicate there is a small chance of very high climate sensitivity (e.g. Andronova and Schlesinger 2001; Murphy et al. 2004; Forest et al. 2002, 2006; see also IPCC, 2007).

The observational weightings usually narrow the width of the climate sensitivity pdf, leading to narrower 5-95% confidence limits, because they contribute information as to which are the most successful predictions. However, the pdf is also influenced by experimental choices: whether it is a 'climate sensitivity' or a 'perturbed physics' ensemble, and the ranges and sampling of input parameters (Forest et al., 2002, 2006; Frame et al., 2005; Rougier, 2007). This is a natural outcome of the Bayesian framework of the ensemble approach. Probabilistic results in the Bayesian sense are defined according to the knowledge available for the analysis: not only the observational data but also the set of hypotheses embodied in the prior distribution (Rougier, 2007).

Ensembles must be large for the results to be statistically sound, and most advanced climate models use substantial amounts of computing time so it may be necessary to use interpolation methods to fill the gaps between a limited number of ensemble simulations (Murphy et al. 2004; Knutti et al. 2006). Despite these various difficulties, the ensemble methodology has been the most significant advance in obtaining probabilistic estimates of climate sensitivity. Uncertainty ranges are more rigorously defined, and the causes of uncertainty are better identified (e.g. Webb et al., 2006).

However, the ensemble approach has not resulted in narrower ranges of climate sensitivity estimates (Figure 2). Many estimates have a width of about 3°C, some much larger, with the upper limit particularly poorly-constrained, and many estimates with the same confidence limits are quite different. This is due both to problems inherent in using the historic period as a reference, which include the small climate signal and uncertainties in the forcings and ocean heat uptake, and to problems inherent in the ensemble approach, which include the sensitivity to experimental choices, the uncertainty associated with interpolation between members of a small ensemble (if used), and the uncertainty inherent in the model itself. The effect of these choices and uncertainties are difficult to quantify, and in some studies no attempt is made to do so.

**Using the palaeorecord**

The geological record includes times when the change in forcing and the climate response were large compared with the changes of the recent past or those expected as a result of doubling $CO_2$. The pre-instrumental or palaeo-record thus provides a strong test of how well we understand and can model different climates, and an opportunity to estimate climate sensitivity in radically different climates. For the most recent glacial-interglacial cycles, the ice core record (e.g. EPICA community members, 2004) provides direct information on global atmospheric GHG concentrations and isotopic measurements of Antarctic surface temperatures. Climates in other eras and in other regions are reconstructed from chemical, isotopic, sedimentological or biological data which reflect the response of these 'sensors' to climate change. However, our knowledge of the climate response to changes in forcing on palaeo-timescales is necessarily less precise than in modern climates, because of the nature of the records and the patchiness of the spatial coverage.

*Inferring climate sensitivity directly from palaeodata*

Just as for modern climates, inferring climate sensitivity from palaeodata records requires estimates of the forcings, the temperature response and the heat uptake by the ocean. However, the rate of ocean heat uptake is usually treated as negligible as it is assumed the ocean is at, or close to, equilibrium, so only $\Delta T_{palaeo}$ and $Q_{palaeo}$ are required (Equation 1). Estimates of climate sensitivity have been made based on three geological periods: the Mid-Cretaceous (Hoffert and Covey, 1992; Barron, 1993), the early Eocene (Covey et al., 1996) and the Last Glacial Maximum (Hoffert and Covey, 1992; Hansen et al., 1993).

The Mid-Cretaceous, about 100 million years ago, was a warm period with atmospheric $CO_2$ concentrations about 2-6 times greater than the present day (Covey et al., 1996). Climate sensitivity estimates based on the Mid-Cretaceous (Table 2; Figure 3) include 3.8±2.0°C (Barron, 1993) and 2.5±1.2°C (Hoffert and Covey, 1992), and the difference between these largely reflects uncertainties in the climate forcing. Barron (1993) considers only $CO_2$ forcing, while Hoffert and Covey (1992) also include forcing due to albedo changes which approximately doubles the total forcing. Their global mean temperature estimates are similar: Hoffert and Covey (1992) obtain the value 9±2°C, using latitudinal reconstructions by Barron (1983) from a variety of data including foraminifera, coral reefs and the lack of permanent ice, while Barron (1993) uses 7±2°C, an updated estimate. The larger forcing thus results in the lower estimate of climate sensitivity (Covey et al., 1996). The early Eocene, 55 million years ago, was also a warm period with atmospheric $CO_2$ concentrations about 2-6 times greater than the preindustrial era. Covey et al. (1996) estimate a climate sensitivity of between 0.7-2.5°C, where the low end of the range corresponds to the highest estimates of atmospheric $CO_2$ concentrations.

The Last Glacial Maximum (LGM: ca 21 000 years ago), corresponding to the global but not necessarily local maximum in ice volume, is characterised by large northern-hemisphere ice sheets, low sea levels, low levels of GHGs and high atmospheric levels of dust (Peltier et al. 2004; Monnin et al., 2001; Dallenbach et al., 2000; Fluckiger et al., 1999; Claquin et al., 2003). The difference in forcing from the present is large and reasonably well-known. Climate sensitivity estimates based on the LGM include 2±0.5ºC (Hoffert and Covey, 1992) and 3±1ºC (Hansen et al., 1993). Hoffert and Covey estimate global mean temperature from the ice core data and the gridded sea surface temperature (SST) reconstructions by the Climate: Long range Investigation Mappings and Predictions (CLIMAP) project. The gridded CLIMAP data are known to be too warm (Broccoli and Marciniak, 1996; Kucera et al., 2005). Hoffert and Covey (1992) also assume that cooling over land is the same as cooling over the ocean at the same latitudes, though this was not the case (Farrera et al., 1999). So their estimate of LGM global mean cooling, -3±0.6°C, is probably too warm and thus yields too small a climate sensitivity (Covey et al., 1996). Hansen et al. (1993) assume LGM global mean cooling is -5±1°C which results in the larger estimate of climate sensitivity.

Climate sensitivity has also been estimated from cyclical climate changes. The Quaternary era, the last 2.6 million years, has been characterised by the occurrence of periodic ice age cycles, each lasting about 100 000 years, which are driven by changes in the Earth's orbit. The relationship between $CO_2$ levels and temperature during these cycles is complicated, because $CO_2$ is both a forcing and a feedback, and the relative timescales of the records are difficult to calibrate. Genthon et al. (1987) and Lorius et al. (1990) use linear regression to analyse the relationship between $CO_2$ and temperature during the last 160 000 years. However, they obtain very different results: 5.4–15.0°C (Genthon et al., 1987) and 3–4°C (Lorius et al., 1990). Lorius et al. (1990) obtain a lower estimate because they assume a smaller Antarctic temperature change (5°C compared with 9°C; the latest estimate is around 11°C: Jouzel et al., 2003), and attribute a larger proportion of the change to $CO_2$ forcing (40% of the temperature response compared with 20%) than Genthon et al. (1987). Lea (2004) takes advantage of a new timescale calibration between the ice core and marine records to analyse a longer period, 360 000 years, and estimates that the 'tropical climate sensitivity' is 5.1±0.8°C. This is extremely high: global sensitivity is expected to be larger than tropical sensitivity, due to the large positive feedback of the polar ice sheets.

Inferences of climate sensitivity based on palaeodata yield estimates that are broadly similar to those obtained from consideration of the instrumental era (Figure 2, Figure 3, Table 1, Table 2; uncertainty limits of 1σ correspond to 68%CL and 2σ to 95%CL). However, these palaeodata-based estimates must be regarded with caution. The forcings are very uncertain (including the division between forcings and feedbacks, and the $CO_2$ component of the total), particularly for eras before the ice core record. The global mean temperature is also very uncertain, as it has been estimated from relatively few data points. And in early periods such as the Eocene and Mid-Cretaceous the geography was radically different. These issues lead to a wide range of estimates for the feedback parameter from each era.

So these estimates based on earlier periods may not be trustworthy, and these studies may only be useful because the changes are generally thought to be due to higher $CO_2$ levels than we will experience in the next century. The following section describes a better approach: using a climate model to simulate global mean temperature and using palaeodata to evaluate the model.

*Constraining model estimates of climate sensitivity with palaeodata*

An alternative to the palaeodata-based approach is to validate palaeoclimate simulations using palaeodata and use the validated models to estimate climate sensitivity. Model validations may take the form of qualitative comparisons to mapped palaeodata or quantitative comparisons to reconstructed temperature changes.

Several eras have been used for validation: the Mid-Cretaceous (Barron et al., 1995), the LGM (Manabe and Broccoli, 1985; Hewitt and Mitchell, 1997; Broccoli, 2000; Hewitt et al., 2001), and the Maunder Minimum (1645-1715), during which sunspots were rare and insolation was low (Rind et al., 2004). After validation, climate sensitivity is estimated either from the palaeoclimate simulation, using $\Delta T_{palaeo}$ and $Q_{palaeo}$ (Barron et al., 1995; Rind et al., 2004) or from a doubled $CO_2$ simulation from the same model, obtaining $\Delta T_{2xCO2}$ directly (Manabe and Broccoli, 1985; Hewitt et al., 2001). One advantage of estimating climate sensitivity from a doubled $CO_2$ simulation is that it does not assume the feedback parameter (Equation 1) is constant in different climate states. Comparisons of LGM and doubled $CO_2$ simulations (e.g. Hewitt and Mitchell, 1997; Broccoli, 2000) indicate that the feedback parameter is probably not constant.

These climate sensitivity estimates are based on palaeoclimate simulations from individual models. However, the Palaeoclimate Modelling Intercomparison Project (PMIP: Braconnot et al., 2007) has shown that even robust responses to changes in forcing vary in magnitude from model to model (Joussaume et al., 1999). Uncertainty in climate sensitivity estimates may therefore be explored by comparing palaeoclimate simulations from different models to each other and to palaeodata. Four of the PMIP models have estimates of climate sensitivity in the range 2.1-3.9°C (Crucifix, 2006). The models have similar estimates of the feedback parameter at the LGM ($\lambda_{LGM}$), but differ in estimates of the feedback parameter in the doubled $CO_2$ climate ($\lambda_{2xCO2}$) and do not agree whether $\lambda_{2xCO2}$ is smaller or larger than $\lambda_{LGM}$. These differences are largely due to the different behaviour of the cloud feedback in each model (Crucifix, 2006). In this study, the limited amount of LGM data used in evaluating the simulations (regional temperature averages over Antarctica, Greenland and the tropical oceans) do not distinguish which is the best model at simulating the LGM and thus most likely to be successful at estimating climate sensitivity.

Perturbed physics ensembles have been used to test the impact of model uncertainties on climate sensitivity and LGM climate simulations (Annan et al., 2005; Schneider von Deimling et al., 2006). In both studies, a regional temperature change in LGM simulations ($\Delta T_{LGM}^{regional}$) is plotted as a function of global temperature change in doubled $CO_2$ simulations ($\Delta T_{2xCO2}$), with one point for each ensemble member (Figure 4). Reconstructions of the LGM temperature change (SST changes averaged over the tropics or other regions) provide numerical constraints on the ensemble estimates of climate sensitivity. Schneider von Deimling et al. (2006) use the 1σ limits of the palaeodata to define the limits of acceptance in the ensemble, while Annan et al. (2005) weight the ensemble members by assuming the palaeodata uncertainties have a Gaussian distribution. The ensemble methodology has the advantage that there is no need to quantify the feedback parameter for either the palaeo- or doubled $CO_2$ climates.

In the published studies, the results differ in part due to experimental choices: they use different forcings, analyse different regions, compare to different temperature reconstructions, and furthermore Annan et al. (2005) estimate only the upper limit of climate sensitivity because their ensemble has few members with low climate sensitivity. However, the results also differ due to differences in the models. The model used by Schneider von Deimling et al. (2006) has a strong linear correlation between $\Delta T_{LGM}^{regional}$ and $\Delta T_{2xCO2}$, which may reflect the simple structure of their 'intermediate complexity' model. The model used by Annan et al. (2005) is a complex GCM, albeit with low resolution and a simplified ocean, and it has a broader, more scattered relationship: in other words, perturbing the physics parameters does not affect the LGM and $2xCO_2$ climate simulations equally. When the models are compared using the same forcings and regional temperature, and compared with the same palaeodata (Figure 4; Annan, private communication; Schneider von Deimling, private communication), it can be seen that the model differences result in different climate sensitivity estimates.

The relationship between $\Delta T_{LGM}^{regional}$ and $\Delta T_{2xCO2}$ in the model simulations (Figure 4) is a measure of the relationship between $\lambda_{LGM}$ and $\lambda_{2xCO2}$, which differs between the two models. Hargreaves et al. (2007) further analyse the ensemble of Annan et al. (2005) and find that most members predict that $\lambda_{LGM}$ is larger than $\lambda_{2xCO2}$ but about one fifth predict the opposite. This result, along with that of Crucifix (2006), illustrates how much uncertainty remains in modelling the response of the Earth to different forcings.

**Combining instrumental and palaeorecord constraints**

Palaeoclimate estimates of climate sensitivity are useful because they examine large climate changes but suffer from increased uncertainties in the climate and forcing estimates, while modern climate estimates have the reverse characteristics. Recent studies (Hegerl et al., 2006; Annan and Hargreaves, 2006) have therefore combined the two types of constraint. Hegerl et al. (2006) constrain a climate sensitivity ensemble of the last 700 years using both the instrumental period (Frame et al., 2005) and palaeodata, while Annan and Hargreaves (2006) combine results from $20^{th}$ century warming, volcanic cooling and LGM cooling. Both studies narrow the range in the estimated climate sensitivity. However, these studies raise a number of issues about combining information from different experiments, including whether it is appropriate to assume that the feedback parameter is constant for different types of forcing (volcanic sulphate aerosols and GHGs affect the climate in very different ways), how to combine qualitative and quantitative estimates of climate, and how best to deal with subjective choices that must be made about which estimates to combine.

**Discussion**

Attempts to estimate climate sensitivity using palaeodata produce a range of estimates just as those obtained using modern observations do (Figure 2; Figure 3). Thus, although the palaeo-record offers advantages over the modern observations because the climate change signal is large compared to the short-term natural variability, attempts to use this record have so far done little to constrain the uncertainties in estimating climate sensitivity. To some extent, and especially for earlier periods in the Earth's history, this reflects the large uncertainties in specifying the change in forcing. A more important issue, however, is the limited use that has been made of palaeodata to constrain the simulations.

Most of the attempts to constrain climate sensitivity using palaeodata are based on regional averages of point-based climate reconstructions (e.g. Crucifix, 2007; Annan et al., 2005; Schneider von Deimling et al., 2006; Hegerl et al., 2006). However, and especially as one goes back further in time, the number of sites for which quantitative reconstructions have been made becomes more limited. There are large uncertainties involved in averaging a limited amount of point data together to create a regional average, and this is especially true when the distribution of the point data is irregular and when there may be no data available for some areas. In these circumstances, spatially-explicit comparisons of simulated climates with palaeoclimate reconstructions provide a stronger assessment of the ability of a model to reproduce palaeoclimates than comparisons based on regional averages. Coupled ocean-atmosphere models may simulate very different spatial patterns of climatic variables such as SST (Figure 5). It is highly plausible that estimates of climate sensitivity will be improved by taking the patterns of climate change into account: part of the uncertainty in climate sensitivity is related to cloud cover, and especially the formulation of stratocumulus, which is strongly influenced by spatial patterns in SST (Webb et al., 2006).

Spatially-explicit reconstructions of climatic variables, based on a variety of different palaeoenvironmental records (including pollen-based vegetation reconstructions, tree-rings, isotopic and noble gas measurements from the terrestrial realm, and biological and chemical proxies from the marine realm), exist for epochs such as the last interglacial, the LGM and intervals during the last glacial-interglacial cycle (Bartlein et al., 1986; LIGA Members, 1991; Cheddadi et al., 1997; Farrera et al., 1999; de Vernal et al., 2000; Peyron et al., 1998, 2000). Although these data sets have been routinely used in model evaluation exercises (e.g. Joussaume et al., 1999; Pinot et al., 1999; Peyron et al., 2006), only a limited subset of the information has been used in attempts to quantify climate sensitivity. This is partly a reflection of the failure of the community to make up-to-date reconstructions readily available. But it also reflects serious concerns about climate reconstructions due to uncertainties in, for example, the direct role of $CO_2$ changes through plant physiology in influencing terrestrial biology (Cowling and Sykes, 1999; Harrison and Prentice, 2003). Forward-modelling techniques which take into account the effects of non-climatic parameters on vegetation changes have been developed (Prentice et al., 2007), but have yet to be applied to continental-scale terrestrial data sets. Again, use of these reconstructions will provide a stronger constraint on the ability of a model to reproduce observed palaeoclimate changes. Finally, there are many sorts of palaeoenvironmental data that reflect changes in climate but which do not yield climate reconstructions. Large-scale syntheses of changes in climate sensors such as, for example, vegetation cover, the extent of lakes, snowline elevation, mineral-dust deposition and charcoal records of palaeofires (Hoelzmann et al., 1998; Prentice et al., 2000; Kohfeld and Harrison, 2001, 2003; Mark et al., 2005; Power et al., submitted) also provide information about the nature of palaeoclimate conditions at a given time. With the advent of more complex climate models which explicitly simulate vegetation, fire disturbance, land-surface hydrology and biogeochemical cycles (including the dust cycle), these data could also be used to provide better constraints on model-based estimates of climate sensitivity. Comparing simulations of climate sensors (such as vegetation) directly with palaeodata, rather than using statistical reconstructions of climate variables, uses as much information from the palaeodata as possible and this increases confidence in the constraints.

Although the use of palaeoclimate targets has not reduced the uncertainty associated with estimates of climate sensitivity, these studies have reinforced our understanding that climate sensitivity is affected by the type of forcing (Hansen et al., 1997; Wigley, 1994; Joshi et al., 2003). Estimates of the feedback parameter based on other forcings than $CO_2$, such as volcanic forcing, will not necessarily yield similar results and could lead to an arbitrary narrowing of the range for climate sensitivity (see, for example, Hegerl et al., 2006). Comparison of LGM, Eocene, Mid-Cretaceous and modern studies suggest that the feedback parameter may also be affected by the size and direction of $CO_2$ forcing.

**Summary**

Current estimates of climate sensitivity based on modern climate data, either used alone or in conjunction with models, are largely in the range 1.5-5°C. Despite improvements in methodology and progress in quantifying the uncertainty, this range has changed little since the first estimates of climate sensitivity were made. This is in large part because the global temperature changes during the 19$^{th}$ and 20$^{th}$ centuries are small, and there are uncertainties associated with some components of the forcing (e.g. insolation, sulphate aerosol forcing) and with the rate of ocean heat uptake.

Past geological periods offer the opportunity to examine climate sensitivity when the climate was radically different from the present, and thus the signal-to-noise ratio is much improved compared to the instrumental period. However, attempts to constrain climate sensitivity using palaeoclimate data have not yet succeeded in substantially reducing the range of estimates. This is in part because the uncertainties associated with the forcings are larger than the uncertainties associated with recent changes in forcing, but also reflects a less-than-optimal use of the palaeoclimate data. Uncertainties also arise from assuming that the behaviour of feedbacks are the same in palaeoclimates as in doubled $CO_2$ climates, and in some studies due to the calibration of timescales (e.g. in records of glacial cycles) and the assumption that the climate is at equilibrium (e.g. at the LGM).

Nevertheless, work to date suggests that palaeodata can help to improve constraints on climate sensitivity. We suggest that a strategy to derive more robust constraints on climate sensitivity should involve:

1. Comparing model simulations to spatially-located data rather than regional averages
2. Using more of the available palaeodata syntheses and palaeoclimate reconstructions
3. Comparing model simulations of climate sensors directly with palaeodata, rather than climate reconstructions
4. Creating new palaeodata synthesis and reconstructions, and working towards well-defined confidence levels for these
5. Extending the current range of palaeoclimate model ensembles
6. Combining constraints from different eras.

*Acknowledgements*

This paper arises out of work being done within the QUEST-funded PalaeoQUMP project. Thanks to PalaeoQUMP project members for useful discussions and to Yan Zhao for producing Figure 5 from PMIP 2 model data (more information about PMIP 2 can be found at http://pmip2.lsce.ipsl.fr/pmip2). We thank James Annan and Thomas Schneider von Deimling for providing unpublished results for analysis.

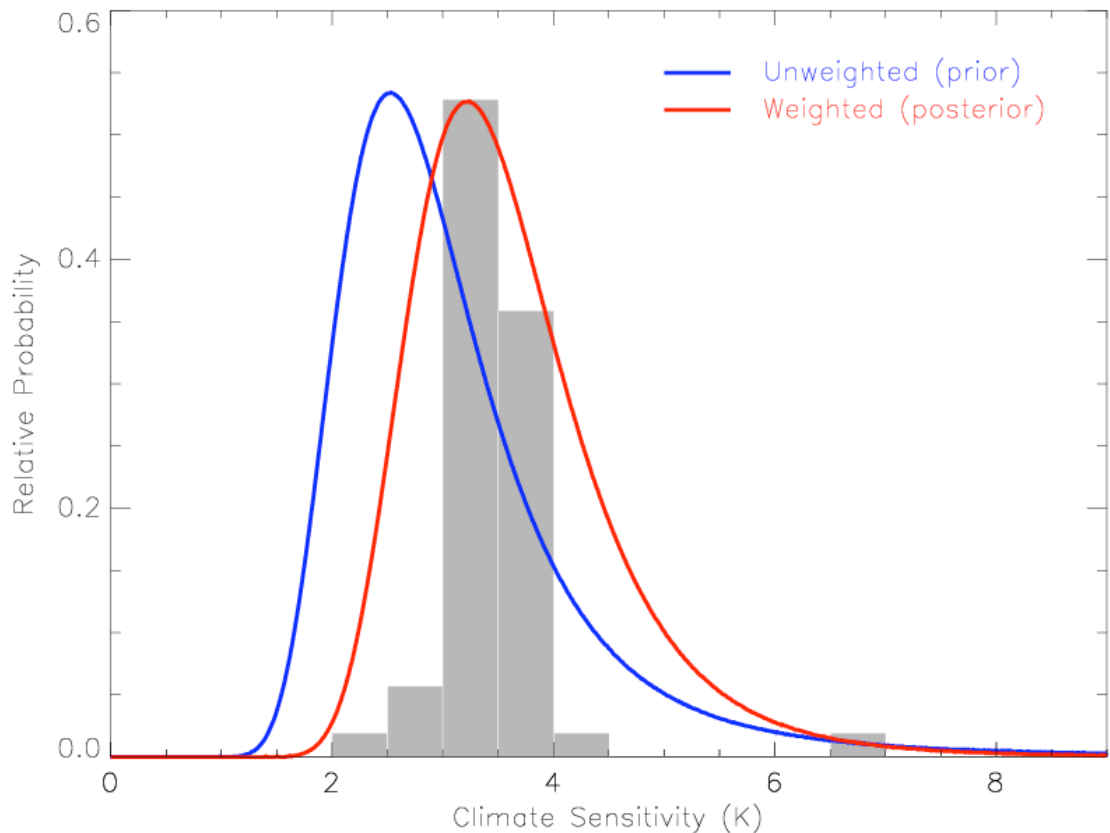

Figure 1. Probability distributions of climate sensitivity (adapted from Murphy et al., 2004). These were obtained from a large perturbed physics ensemble (grey histogram), using linear interpolation to predict the results from additional parameter combinations. The pdfs are shown with (red) and without (blue) weighting according to the agreement between model versions and present day climate observations.

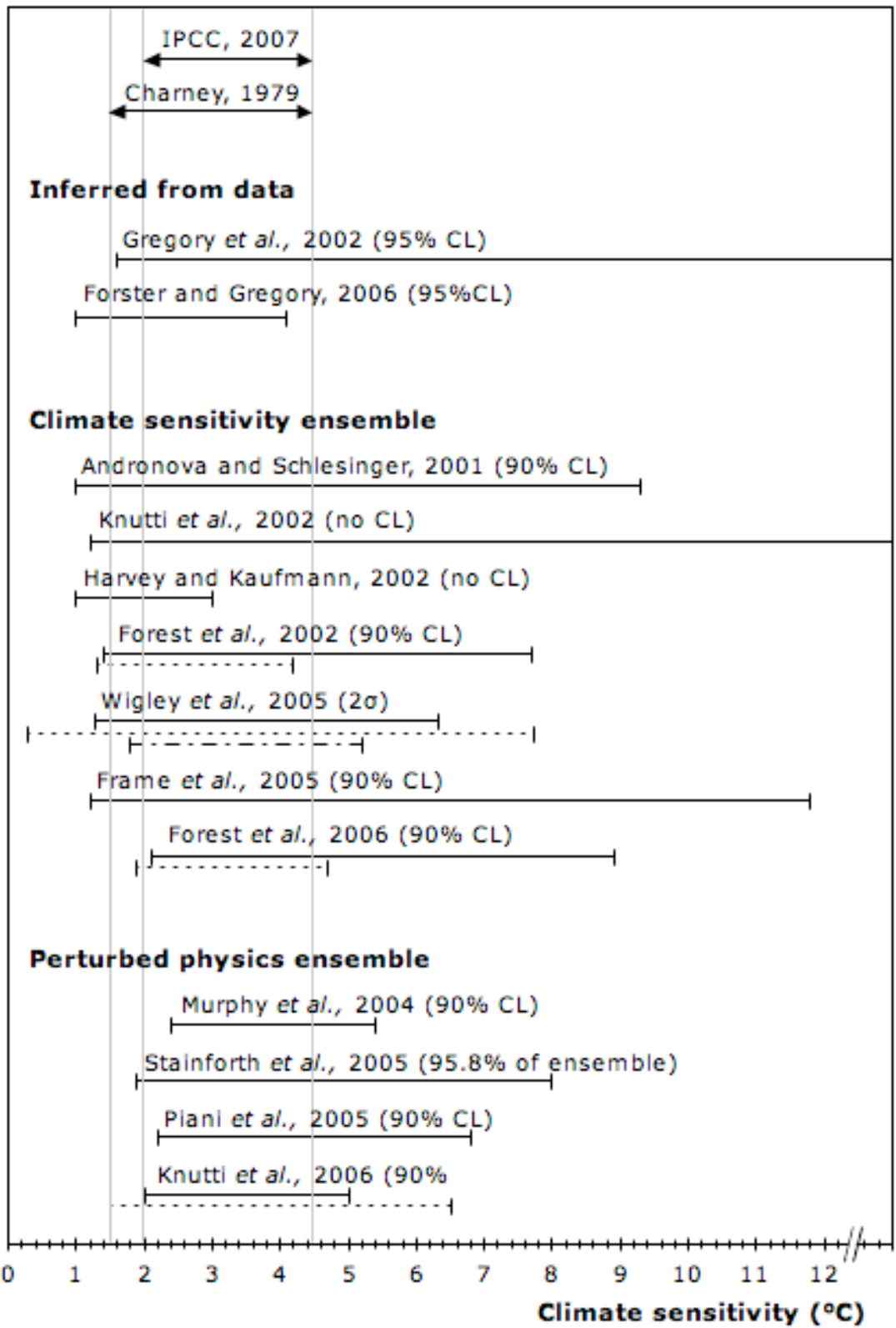

Figure 2. Climate sensitivity estimates obtained using modern climate data as a constraint, divided into three groups: inference of climate sensitivity directly from data; ensemble studies in which climate sensitivity is varied; and ensemble studies in which physics process parameters are varied. Wigley et al. (2005) base their estimates from three different volcanic eruptions (Table 1). The narrower ranges of Forest et al. (2002; 2006) include additional specifications for the values of the ensemble input

parameters. Confidence limits, or other definitions of the estimate, are given. The vertical shaded bands indicate the IPCC 2007 (2-4.5ºC) and NAS (1.5-4.5ºC) and ranges (see text).

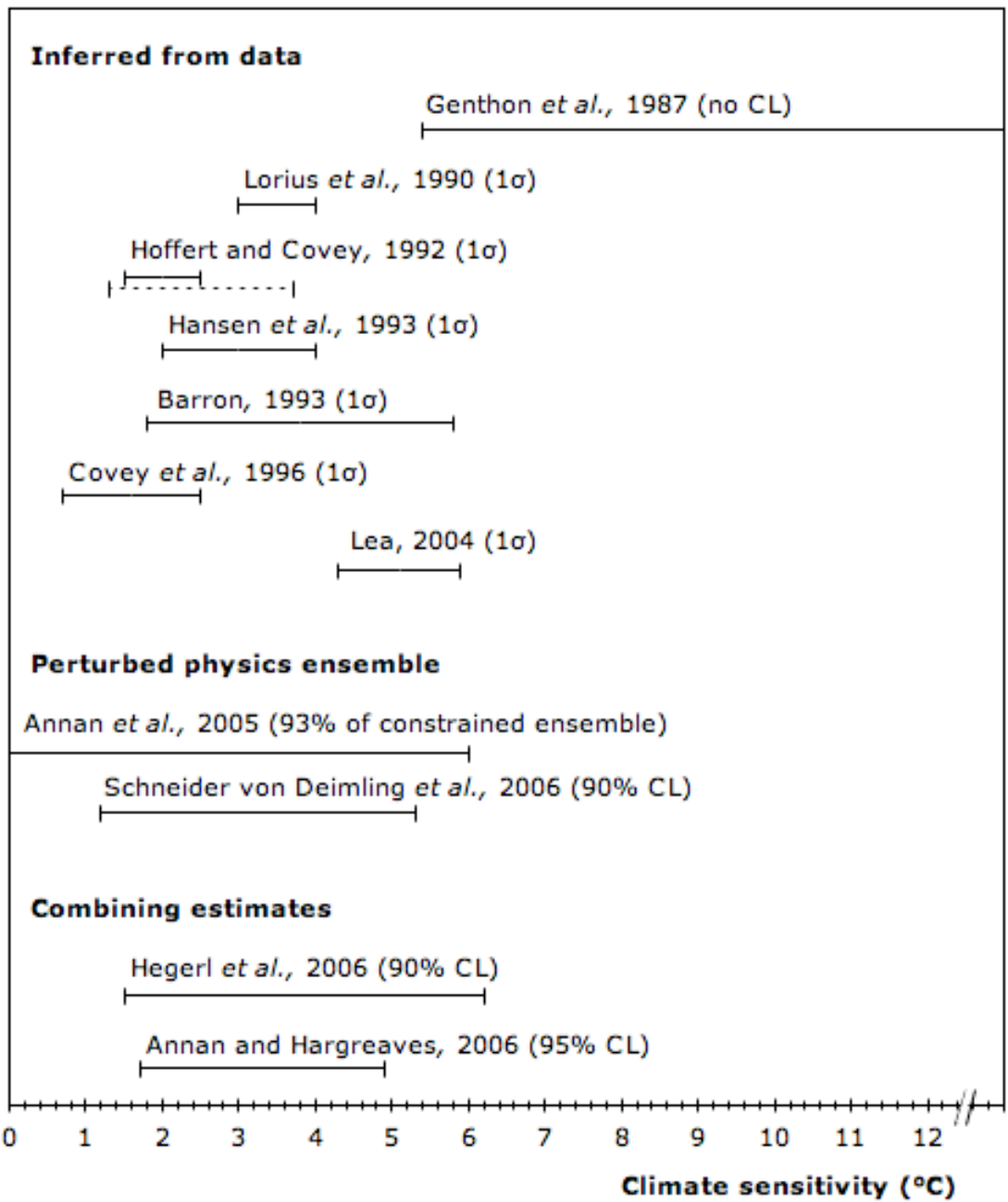

Figure 3. Climate sensitivity estimates obtained using palaeoclimate data as a constraint, divided into three groups: inference of climate sensitivity directly from palaeodata; ensemble studies in which physics process parameters are varied; and results from combining estimates from different eras. Hoffert and Covey base their estimates on two eras: the LGM (solid line) and Mid-Cretaceous (dashed line).

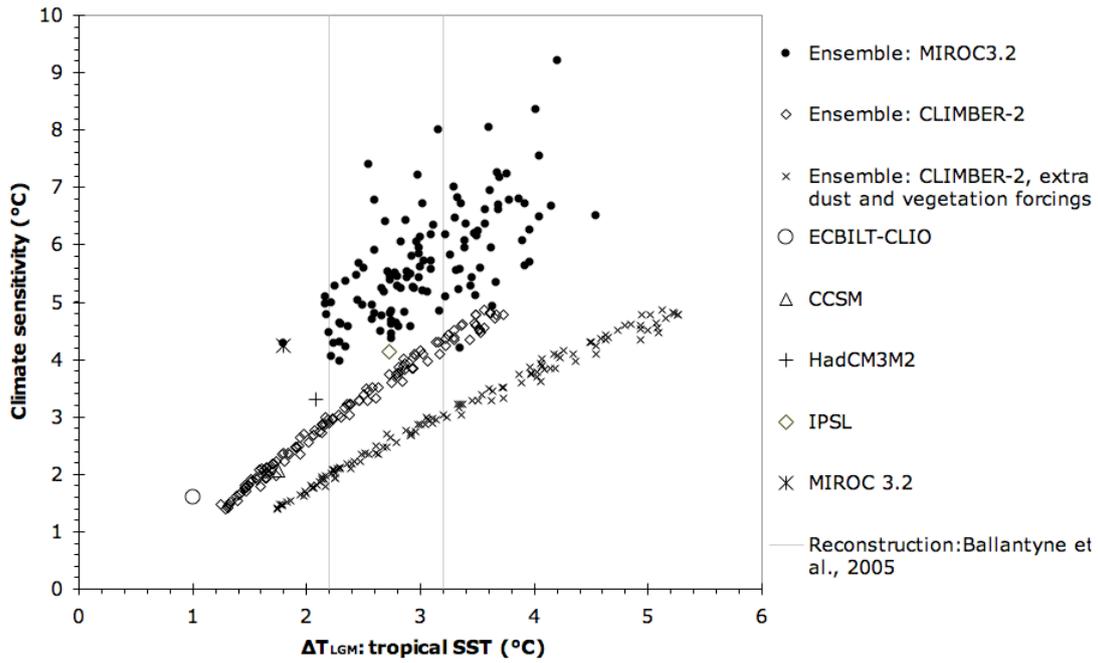

Figure 4. Climate sensitivity as a function of tropical sea surface temperature change between the pre-industrial and the LGM for three model ensembles: the MIROC3.2 model with PMIP LGM boundary conditions (Annan, private communication); the CLIMBER-2 model with PMIP LGM boundary conditions; and CLIMBER-2 with additional dust and vegetation forcings (Schneider von Deimling, private communication). For comparison, five PMIP 2 coupled ocean-atmosphere GCMs are shown (Crucifix 2006; this paper). The MIROC3.2 ensemble uses a simpler version of the model than PMIP 2 (see text). The vertical lines indicate the 1σ limits of reconstructed tropical SST change at the LGM from Ballantyne et al. (2005).

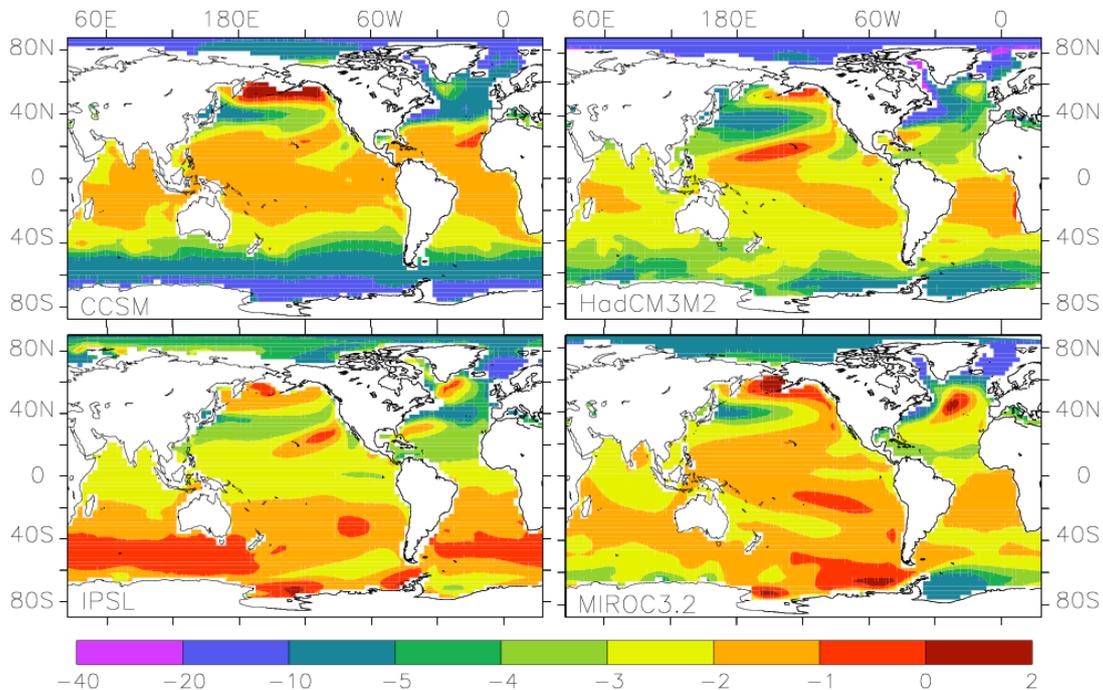

Figure 5. Annual mean sea surface temperature changes between the pre-industrial and the LGM, simulated by four coupled ocean-atmosphere GCMs from PMIP 2 (Braconnot et al., 2007): CCSM, HadCM3M2, IPSL-CM4-V1-MR and MIROC3.2.

| Authors | Instrumental data | Climate sensitivity |
|---|---|---|
| **Inferred from data** | | |
| Gregory et al., 2002 | Ocean and surface air temperature | Lower limit 1.6ºC (95% CL) |
| Forster and Gregory, 2006 | Net flux imbalance at top of the atmosphere; surface air temperature | 1.0–4.1ºC (95% CL) |
| **Model validation** | | |
| Barnett et al., 2001 | Ocean temperature | 2.1ºC consistent with data |
| Yokohata et al., 2005 | Surface air temperature (volcanic cooling) | 4ºC consistent with data; 6.3ºC not consistent |
| **Climate sensitivity ensemble** | | |
| Andronova and Schlesinger, 2001 | Surface air temperature | 1.0–9.3ºC (90% CL) |
| Knutti et al., 2002 | Ocean and surface air temperature | Lower limit 1.2ºC |
| Harvey and Kaufmann, 2002 | Ocean and surface air temperature (inc. volcanic cooling) | 1.0–3.0ºC favored; 2ºC most likely |
| Forest et al., 2002 | Ocean, surface air and upper air temperature | 1.4–7.7 (90% CL) with uniform prior; 1.3–4.2 (90% CL) with expert prior |
| Wigley et al., 2005 | Surface air temperature (volcanic cooling) | Agung: 2.83ºC (2σ limits 1.28–6.32ºC); El Chichón: 1.54ºC (2σ limits 0.30–7.73ºC); Pinatubo: 3.03ºC (2σ limits 1.79–5.21ºC) |
| Frame et al., 2005 | Surface air temperature | 1.2-11.8ºC (90%CL) |
| Forest et al., 2006 | As for Forest et al., 2002 | 2.1–8.9ºC (90% CL) with uniform prior; 1.9–4.7ºC (90% CL) with expert prior |
| **Perturbed physics ensemble** | | |
| Murphy et al., 2004 | Large range of present-day observations | 2.4–5.4ºC (90% CL) |
| Stainforth et al., 2005 | Present-day annual mean temperature, sea level pressure, precipitation and atmosphere-ocean sensible and latent heat flux | 1.9 –11.5ºC unconstrained range of ensemble; (95.8% of ensemble < 8ºC) |
| Piani et al., 2005 | As for Stainforth et al. (2005), plus relative humidity, zonal and meridional winds, outgoing LW and SW radiation | 2.2–6.8ºC (90% CL) |
| Knutti et al., 2006 | Present-day surface temperature seasonal cycle | Lower limit 1.5–2ºC (90% CL); Upper limit 5–6.5ºC (90% CL) |

Table 1. Estimates of climate sensitivity that use modern climate data (post-1850) as a constraint.

| Authors | Era | Climate sensitivity |
|---|---|---|
| **Inferred from data** | | |
| Genthon et al., 1987 | Last 160 000 years | 5.4–15.0°C |
| Lorius et al., 1990 | Last 160 000 years | 3–4°C (1σ) |
| Hoffert and Covey, 1992 | LGM; Mid-Cretaceous | LGM: 2±0.5°C (1σ); Mid-Cretaceous: 2.5±1.2°C (1σ) |
| Hansen et al., 1993 | LGM | 3±1°C (1σ) |
| Barron, 1993 | Mid-Cretaceous | 3.8±2.0°C (1σ) |
| Covey et al., 1996 | Eocene | 1.6±0.9°C (1σ) |
| Lea, 2004 | Last 360 000 years | Tropical sensitivity: 5.1±0.8°C (1σ) |
| **Model validation** | | |
| Manabe and Broccoli, 1985 | LGM | 2.3°C and 4.0°C consistent with data |
| Barron et al., 1995 | Mid-Cretaceous | 3°C |
| Hewitt and Mitchell, 1997 | LGM | 2.9°C reasonably consistent with data |
| Broccoli, 2000 | LGM | 3.2°C |
| Hewitt et al., 2001 | LGM; modern | 3.3°C and 2.8°C consistent with palaeodata; 2.8°C better agreement with modern data |
| Rind et al., 2004 | Maunder Minimum | 4.4°C not consistent with the data; 1.1°C would be consistent |
| Crucifix et al., 2006 | LGM | 2.1–3.9°C (unconstrained model range) |
| **Perturbed physics ensemble** | | |
| Annan et al., 2005 | LGM | Greater than 6°C hard to reconcile with data; greater than 8°C virtually impossible |
| Schneider von Deimling et al., 2006 | LGM | 1.2–5.3°C (90% CL) |
| **Combining constraints** | | |
| Hegerl et al., 2006 | 20th century; last 700 years | 1.5–6.2°C (90% CL) |
| Annan and Hargreaves, 2006 | 20th century; volcanic cooling; LGM cooling | 1.7–4.9°C (95% CL) |

Table 2. Estimates of climate sensitivity that use palaeoclimate data (pre-1850) as a constraint.